# Substrate-Enzyme Interaction in Pig Liver Esterase

D. Hasenpusch[1], D. Möller[1], U.T. Bornscheuer[2], and W. Langel[1]*

[1] Department of Biophysical Chemistry, Institute of Biochemistry, University of Greifswald

[2] Department of Biotechnology & Enzyme Catalysis, Institute of Biochemistry, University of Greifswald

*Corresponding author: W. Langel, Department of Biophysical Chemistry, Institute of Biochemistry, Felix-Hausdorff-Str. 4, 17487 Greifswald, langel@uni-greifswald.de

## Abstract

Force field and first principles molecular dynamics simulations on complexes of pig liver esterase (pig liver isoenzymes and a mutant) and selected substrates (1-phenyl-1-ethyl acetate, 1-phenyl-2-butylacetate, proline-β-naphthylamide and methyl butyrate) are presented. By restrained force field simulations the access of the substrate to the hidden active site was probed. For a few substrates spontaneous access to the active site via a well defined entrance channel was found.

The structure of the tetrahedral intermediate was simulated for several substrates and our previous assignment of GLU 452 instead of GLU 336 was confirmed. It was shown that the active site readily adapts to the embedded substrate involving a varying number of hydrophobic residues in the neighborhood. This puts into question key-lock models for enantioselectivity. Ab initio molecular dynamics showed that the structures we found for the tetrahedral intermediate in force field simulations are consistent with the presumed mechanism of ester cleavage. Product release from the active site as final step of the enzymatic reaction revealed to be very slow and took already more than 20ns for the smallest product, methanol.

# 1 Introduction

Pig Liver Esterase (PLE) is widely used as biocatalyst in organic synthesis for the last 30 years, especially in the resolution of a wide range of substrates including racemic secondary alcohols [1]. The hydrolysis of a broad range of racemic or prostereogenic carboxyl esters proceeds with high to excellent enantioselectivity, and PLE is a very useful enzyme for sophisticated organic synthesis processes. Its isoenzymes show activity both as esterases and amidases with different degrees of selectivity for substrates and enantiomers [2]. The globular serin-hydrolase has α/β-fold and about 60 kDa molecular weight [3]. By recombinant expression from *Pichia pastoris* reproducible synthesis of several isoenzymes is now possible [4]. Recently we have established and validated a sterical model for PLE-1 to -6 and APLE including the structural effects of mutations [5] on the basis of sequential homology to the closely related human carboxyl esterase (hCE) [6]. The hydrolysis of chiral esters by PLE is enantioselective as was shown for a homologous series of six chiral substrates [7].

A small number of mutations close to the active sites of enzymes may be sufficient for significant modification or even reversal of enantioselectivity [8]. Efficient mutations are within 11 Å around the active site, and at higher distances the respective E-values are modified by less than a factor of three. The enantioselectivity of rPLE was studied experimentally by generating a series of nine mutants with finally 17 exchanged residues representing the transition from genuine PLE-1 to recombinant pig intestinal carboxyl esterase (rPICE) [9]. The sequence of this enzyme differs only at 23 positions from PLE-1 corresponding to a homology of 97% [10]. The first mutant in this series, PLE-1a, contains two mutations with respect to PLE-1, V76A and E77G, with distances from the active site as high as 10-14 Å. Still the (*R*)-preference for 1-phenyl-1-ethylacetate (**1**) (Fig **1**) increases from E=8 in PLE-1 to 60 in PLE-1a. These values sensitively depend on the substrate structure. 1-phenyl-2-butylacetate (**2**) differs from substrate **1** only by two additional methylen groups, but shows a clear preference of the (*S*)-enantiomer with E values as high as 55 and 20 in PLE-1 and PLE-1a, respectively.

Methyl butyrate (**4**) as a small achiral substrate is hydrolyzed by PLE isoenzymes -1 to -5 to butyric acid (**5**) and methanol (**6**), PLE-5 being most active. Conversion of proline-β-naphthylamide molecule (**3**) should be slow since the molecule is larger than the other substrates, which might slow down access to the active site. Early studies have shown, however, that PLE-1, the former γ-isoenzyme, converts **3** rather than **4** [3a].

Several attempts have been made towards understanding the mechanism of enantioselectivity for enzymes since this is essential for protein engineering [11] beyond trial and error mutations. The enzymatic hydrolyzed of esters may be dissected into four steps:

(i) The substrate molecule has to access the active centre of the enzyme from the solution. Complicated channels to hidden active sites have been studied in a few examples, such as Cytochrome P450 [12], where access of educts was simulated by steered molecular dynamics (SMD) [13]. From crystal structure data of the PLE analogs hCE and rabbit liver carboxyl esterase (rCE) it was shown that the access of substrate analogs passes between two $\alpha$-helices which correspond to those formed by residues 72-87 and 338-346 in PLE [14]. They form a flexible gate, and the substrate has to pass to the active centre through a channel with a length of about 15 Å.

As mutations with a considerable distance to the active site are efficiently modulating enantioselectivity of PLE, we will consider the possibility that not only the active site but also remote parts of the enzyme, especially the entrance channel, play a major role [15]. An important result of our PLE model was that the mutations, which distinguish isoenzymes with different substrate specifities and enantioselectivities [2b], affect the structure of the α-helices around the entrance channel rather than the active site. In PLE 3 even the helical folding of the entrance is suppressed in part, and the entrance hole is wider than in the other isoenzymes [5].

(ii) The molecule has to fit into enzyme binding pockets in a position enabling the interaction with the catalytic triad consisting of SER 204, HIS 449 and a GLU residue. By analogy to rCE this should be GLU336, but our structure models without substrate were more consistent with GLU 452 [5]. An early model of enantioselectivity was that only one of the two enantiomers matches the rigid geometry of the active site and reacts (“key and lock principle”). More recently the “induced fit” approach assumed that soft pockets around the substrates adapt to their shapes, which may not work equally well for both enantiomers [16]. These two concepts of enantioselectivity preclude that all other processes are fast or non specific, which is the basis for enzyme engineering on the active site [8].

A schematic empirical model for the active site of PLE with a big and a small hydrophobic pocket, $H_L$ and $H_S$, respectively, at opposite positions and two rather polar lateral pockets around the catalytic serine was derived from a series of experiments at chiral substrates [17]. The rear polar pocket $P_B$ stabilizes the free oxygen atom. In hCE and rCE a big flexible and a small rigid pocket, and an “oxyanion hole” were defined corresponding to $H_L$, $H_S$, and $P_B$, respectively.  Amino acids have been assigned to these pockets on the basis of crystal structures [14] with embedded large substrates.

(iii) Ester cleavage inside the active site passes via three elementary steps, starting with bond formation between the $O_\gamma$ of serine in the catalytic triade and the ester carbon atom of the substrate. The enantioselectivity of *Candida antarctica* Lipase B with respect to some amines and secondary alcohols was correlated to the geometry of the tetrahedral intermediates where only one of the two enantiomers fits well to the active site [18]. In a directed evolution study on an enantioselective epoxide hydrolase from *Aspergillus niger* the difference of the distances between the (*R*)- and (*S*)-substrates and a stabilizing ASP residue in the tetrahedral intermediate was correlated with the respective E-value [19]. Another quantitative approach uses force field MD for calculating the potential energies of the (*R*)-and (*S*)-species in the active site by a special docking method. The difference was set equal to the that of free activation enthalpies for the transition states of the enzymatic reaction, $\Delta\Delta G^{\#}$, which is directly related to enantioselectivity [20].

The two crucial steps provide breaking of the bonds inside the ester and to the serine, setting free the alcohol and the acid, respectively. In the mechanism proposed in (Ref [1], p. 69), the alcoholate anion withdraws a proton form the histidine in the active site, and the carboxylic acid is obtained as the acyl rests binds to a hydroxyl group from a water molecule. The remaining proton is restituted to the histidine.

(iv) Slow release of products from the active site would block the next cycle of the enzymatic reaction and the respective processes are similarly relevant for the enzymatic reaction as the access of the educts to the active site. By steered molecular dynamics the egress of metypyron from cytochrome P450 was modeled [21]. Random expulsion molecular dynamics (REMD) was also used for modeling release of products from the active site of the same enzyme but the authors claim that channels for release of products and access of educts are identical [22]. Beyond these studies few information is available on the removal of reaction products from the active site into the solution.

In principle any of these steps may contribute to the observed kinetics, and steps (i) to (iii) obviously can be enantioselective. The release of achiral products (iv) may only depend on the educt enantiomer if the active site fits to different shapes for transition states with (*R*)- and (*S*)-modifications, respectively. No detailed kinetic data for racemate conversion [23] are available so far for well defined isoenzymes [2b, 7], and it is not known if a given modification of their E-values is due to slowing down the reaction of one enantiomer or to accelerating the other one. Thus presently available experimental data do not reveal, if given modifications of enzyme or substrate result in better or worse fits.

The aim of this project is to understand the reaction mechanism and to find factors contributing to the experimentally observed enantioselectivity of the enzyme on the basis of molecular dynamics simulations. Consequently, we set up calculations for all reaction steps. Results will give indications on possible mechanisms rather than lending themselves to quantitative interpretations. We focus on the differences in enantioselectivity of the very similar isoenzymes PLE-1 to -6 and APLE and their mutant PLE-1a for the related chiral esters **1** and **2.** The two substrates **3** and **4** are smaller and bigger than **1** and **2**, respectively and shall reveal the size specificity of the active sites. Most calculations refer to enzyme monomers rather than trimers for reducing the computational efforts.

An important question is if the active site or the entrance channel contributes to the enantioselectivity. On the basis of our structural model we localize the amino acids adjacent to the substrate and elucidate the structure of the pockets. On the other hand, by two methods of molecular dynamics, structural effects on the access of the substrate to the active site are traced. Restrained molecular dynamics forced insertion of the substrate initially positioned outside into the active site and allowed entrance channels to be identified. In some cases we observed spontaneous access of the substrate within the time scale of the force field simulation. The consistence of current models for the enzymatic reaction with our structure was verified by quantum mechanical calculations on the splitting of the ester bond. Finally we traced spontaneous product release by unrestrained molecular dynamics.

# 2 Method

## *Free substrates*

Free substrate molecules **1** – **4** were generated using gaussview [24]. The substrate structures were optimized in Gaussian G98 [24] on HF/6-31G* level. The electrostatic potentials and partial charges for molecules not implemented in the standard AMBER94 force field were calculated according to the RESP procedure [25] using the antechamber program (supporting information). By theory the partial charges of the two enantiomers of each substrate are identical as long as relaxation in a given chemical environment results in mirror symmetric configurations. We calculated charges for arbitrary relaxed structures of the (*R*)- and (*S*)-forms and obtained deviations for the charges for different configurations of less than 0.05e. The original AMBER94 library did not contain ester compounds, and specific parameters were added (supplementary information).

## Mutants and isoenzymes

In addition to the isoenzymes PLE-1 to -6, the first mutant of PLE-1 from [10] with two (V76A, E77G) exchanged amino acids was generated (PLE-1a). According to the numbering scheme from [5], our PLE monomers contain the residues 4 to 534, since three residues at the N-terminus were omitted during homology modeling. PLE contains five cysteine residues which form two **disulfide-bridges** (70-99, and 256-267) [26]. They were open in some calculations, which did not affect the stability of the protein on the time scale of our simulations. Three identical PLE-1 monomers were assembled to form a trimer.

## Force field simulations of the solvated enzyme with substrates

The computational approach was described in detail before [5]. Systems were generated using the graphical interface xleap of the AMBER package (Versions 7 to 10) [27] and visualized using the VMD program [28] and its plug-ins [29]. For applying periodic boundary conditions the enzymes were embedded into orthorhombic boxes with sizes of 72x84x92 $Å^3$ and 85x137x134 $Å^3$ for monomers and trimers, respectively. Spacing of at least 10 Å between enzyme and box walls minimizes the interaction between the molecule and its periodic images. The boxes were filled with discrete TIP3P water molecules. Under physiological conditions the substrates are neutral but the enzymes had total negative charges of 12 (PLE-1 and -2), 13 (PLE-1a) and 14 (PLE-3 to -6 and APLE), respectively, which were compensated by singly positive counter ions (CIO) corresponding to $Na^+$.

Protonation of the histidine ring at the $N_\varepsilon$-atom is the default in the AMBER force field definition. As was laid out in [5] the active site affords protonation of the $N_\delta$ atom for enabling interaction between SER 204 and HIS 449 residues. The latter was redefined as HID with protonated $N_\delta$ in several runs especially for modeling spontaneous access of the substrate to the active site.

For minimization of the energy and for classical molecular dynamics simulations of enzyme systems (time step 1 fs) the sander program with the AMBER94 force field was employed. First the total potential energy of water molecules and counter ions only was minimized, then of enzyme and substrate only, and finally of the whole system. In an MD run at constant pressure and temperature (**NpT**) for 20 ps, the cell size was adjusted attaining optimum water density. In further short MD runs the enzyme and finally the whole system were relaxed. After initialization the systems were subjected to specific molecular dynamic calculations typically scanning 5 to 30 ns.

## Formation of the Michaelis-Menten complex by restrained molecular dynamics

For restrained molecular dynamics, the substrate was positioned outside of the enzyme and pulled into it by setting and shortening a harmonic restraint with a typical force constant k=14 N/m to the distances between various substrate C-atoms and the $O_\gamma$ of SER 204. The initial set point of the restraint around 39 Å was continuously reduced to 3 Å with velocities between 1 and 4 m/s, typically 2 m/s. Due to friction the actual distance s between the respective C and O atoms is in general back lagging with respect to the set point $s_0$. The force F of the restraint on the substrate molecule is calculated from Hooke's law as $F=k*(s-s_0)$ and plotted over s yielding force distance curves. Reducing k to values between 7 and 12.5 did not have significant influence on the plots. In general we use a starting temperature of 300 K which is controlled in some runs. We changed the controlled temperature in some runs to 400, 500 and 600 K. Simulations for 1.8 ns at these elevated temperatures do not result in denaturation of extended proteins, but in an acceleration of local processes [30].

## Tetrahedral intermediate

In the tetrahedral intermediates the carbonyl C-atom of the ester substrates is covalently bonded to the atom $O_\gamma$ of SER 204. Eight starting structures for the (*R*)- and (*S*)-enantiomers of substrates **1** and **2** in PLE-1 and PLE-1a were generated by manually moving the respective molecules into the active sites and adding the bond using the xleap program. Switching to another substrate is achieved by erasing all but the carbonyl C-atom of the previous one and changing the respective molecule name in the pdb file of the system. Xleap adds missing atoms on the basis of the respective library. Occasionally, some close contacts near the covalent bond between the new substrate and the enzyme had to be corrected manually. Before bonding the substrate to the enzyme, a proton transfer from serine to histidine occurs, creating a deprotonated SER 204 and a doubly protonated HIS 449 residue. The respective histidine (HIP) is taken from AMBER force field amino acid libraries, and an additional residue SEO was derived from standard serine by omitting the proton $H_\gamma$.

The charges of the substrate (supplementary information) as bonded to the $O_\gamma$ of SER 204 were calculated from the electrostatic potential of subsets of atoms, which contained the ester and additional essential amino acids for the enzymatic reaction beyond SER 204. The subset structures were taken from the optimized enzyme without further optimization, and the total charges were set to -1 accounting for the serine deprotonation. The subset size was chosen such that the calculated charges for the included SER 204 closely reproduced those of the predefined SER residue of the force field. We first optimized the subset with substrate (*S*)-**1** yielding charges for serine which coincide within 0.05 e with those of the predefined residue, and then repeated the calculation with substrate (*S*)-**6**. The optimized subsets contained an acetyl group at the N-terminal of SER 204 mimicking a peptide bond. At the C-terminal a glycine residue was added instead of ALA 205, being part of the oxyanion hole, which stabilizes the free $O^-$ of the substrate in the tetrahedral intermediate (s. below). For blocking the terminal carboxyl group the OH was replaced by $NH_2$. In addition to this peptide the side chains of the doubly protonated HIS 449 and the deprotonated GLU 336 were included. As a second part of the oxyanion hole, acetamide mimics a blocked peptide bond between GLY 126 and LEU 127.

## *Enzymatic reaction: Quantum mechanics*

For first principles simulations the Car-Parrinello molecular dynamics package cpmd (3.9.1) [31] with Vanderbilt ultrasoft pseudopotentials [32] and the Perdew-Becke generalized gradient approximation was used [33]. Time step and simulation time were 0.17 fs and 0.8 ps, respectively, and the box size was 20x20x15 Å.

For combined first principles (quantum mechanical) and force field (molecular mechanics) simulations (QM/MM), the extensions of cpmd 3.11.1 und 3.13.2 by the GROMOS force field [34] were set up with BLYP exchange-correlation functional and Goedecker pseudo potentials [35]. Time step was 0.085 fs and total simulation times extended to 2-4 ps. Two cells were defined for the quantum mechanical and the molecular mechanics parts of the system with sizes of 21.2x21.2x21.2 $Å^3$ and 72x84x92 $Å^3$, respectively.

In this model quantum mechanically modeled side chains of the catalytic triad are chemically connected to a backbone described by molecular mechanics. As bonds between atoms in the different models are not defined, the respective $C_\alpha$ and $C_\beta$ carbon atoms were saturated by an MM and a QM hydrogen atom, respectively. The positions of all four atoms were fixed, the three distances ($C_\alpha$-$C_\beta$, $C_\alpha$-$H_{MM}$, and $C_\beta$-$H_{QM}$) having standard values. Simulations on ethane demonstrated that the torsions of the two methyl groups are correctly reproduced by this approach. For the protein simulation it implied that the $C_\alpha$ and $C_\beta$ atoms of the catalytic triad were fixed in space.

# 3 Results and Discussion

## *Access channels to the active site*

### *Restraint molecular dynamics*

We performed restrained classical MD calculations for tracing the access channels of the substrates (*R/S*)-**1** and (*R/S*)-**2** to the active sites of PLE-1 and PLE-1a resulting in Michaelis-Menten complexes. In contrast to manual docking the results should be independent of the initial positioning of the substrate and will indicate its preferred configuration in the active site. Initial substrate positions were chosen by visual inspection of van der Waals surfaces. The two residues, to which the substrates dock, were partly visible from outside through holes indicating pathways, which are too small for straightforward access, but afford relaxation of the amino acids around the entrance. The substrates were manually positioned close to these holes.

During most pulling trajectories the substrate reached the active site with a final distances between ester C and $O_\gamma$ of SER 204 of 3 to 4 Å. For different pulling points in the substrate with the same initial configuration very similar channels were found. Inspection of the trajectories shows that the pathway is not a straight line from starting to final point but follows openings in the protein structure. In a few runs the substrate was stuck at the entrance of the enzyme. The restraint then resulted in distortion of the active sites and an increase of the distances between $O_\gamma$ SER 204 and $N_\delta$ HIS 449. In two runs the substrates (*R*)-**1** and (*S*)-**1** pass through the two entrance helices (**Fig 2**) similar as for spontaneous access (s. below), whereas most channels for restrained access are bypassing them. Along two of these channels outside the entrance helices five intermediate substrate positions each were chosen as starting configurations for free MD runs. The substrate molecules close to the surface left the enzyme and returned to the solution, and further inside they were stuck making it unlikely that spontaneous access is possible this way.

Substrate **3** was positioned around PLE-1 for a number of calculations on restrained access to the active site. Even though the initial substrate positions were spread out on a half circle above the soft part of the enzyme, all channels passed through the entrance helices. Obviously only small molecules can reach the active site on random ways through the protein, but for larger substrates the barriers on these way are too high.

### *Force distance curves*

The forces along each channel fluctuated and typically increased close to the active sites. Peak values were always below 2 nN, typically around 1 nN (**Fig 3**). Runs from identical starting positions with controlled temperatures of 300-600 K resulted in very similar channels. Resistance to pulling decreases with increasing temperature indicating enhanced flexibility of the enzyme structure. At 600 K random thermal motion of the amino acids around the entrance channel seems to superpose noise to the force distance curve, whereas the forces for widening the channel are probably reduced by thermal softening to values below noise level. In [5] we showed that the PLE consists of hard and soft parts being distinguished by different values of root mean square displacements during free molecular dynamics. The pathways found here are passing through soft parts of the enzyme and no way was found through the stiff β-sheets.

From the force distance curves the access to the active site is interpreted as a superposition of viscous flow and hopping over triangular barriers with heights of 0.2 to 0.6 nN and lengths of about 5-10 Å, corresponding to energies around 0.3 to 2 eV (**Fig 3**). Viscosity usually slowly increases on the way into the enzyme, and positions and sizes of the barriers are subject to thermal fluctuations similar to jump diffusion in a liquid. Significant forces are typically found on the last 20 to 30 Å of the way into the active site. This long distance indicates that the substrate experiences friction on the surface of the enzyme before entering it. On the way through the two entrance helices the path length is reduced to 12 Å.

In spite of the large scatter of the data, the force seems to increase with increasing pulling velocity. Forces of the order of magnitude as observed here can be mimicked by pulling the molecule through a liquid with a viscosity around 0.03 Pa·s, as for viscose oil. This estimate uses the Stokes formula and a radius of 4.2 Å for substrate **1** with a spherical shape and a density of 0.9 g/cm$^3$. This low effective viscosity is consistent with our claim that the active site adapts its shape according to the substrate by an induced fit.

### *Spontaneous access to the active site: Unrestrained MD of enzymes with substrates*

Substrates will spontaneously enter the enzyme only from specific positions on its surface. In a real system they are sampled by Brownian motion over times long as compared to real times of MD runs. Here we have to initiate this process by placing substrate molecules close to presumed entries, and by using various starting configurations. In one series of calculations we addressed the difficulty of finding appropriate starting positions by placing an excessive amount of substrate (*R*)-**1** (20976 molecules), (*S*)-**1** (9312 molecules), (*R*)-**2** (9090 molecules), and (*S*)-**2** (5550 molecules) around the trimer. During these runs substrate molecules did not enter the active site, probably due to mutual interference, but several of them passed between the entrance helices, where would be the starting positions for the entrance channel. All these molecules first had contact with the residue HIS 288, which is part of a loop on the surface pointing into the solution, and we speculate that this residue might trigger the access to the active site.

More realistically, systems with a single molecule of each of the six substrate species ((*R/S*)-**1** and (*R/S*)-**2**, **3** and **4**) and PLE-1 were set up, and then duplicated mutating PLE-1 to PLE-1a. In these runs the $N_\varepsilon$ HIS 449 was protonated. During the first nanoseconds of the simulation time the mutated enzyme PLE-1a undergoes structural relaxation as a consequence of the two mutations. Spontaneous access of **4** to the active sites of PLE-1 and PLE-1a was observed within 1 ns.

Relaxed structures for the seven isoenzymes PLE-1 to -6 and APLE with HIS 449 protonated at $N_\delta$ were taken from [5], and four molecules of each of the five substrates (*R/S*)-**1** (*R/S*)-**2** and of **4** were positioned close to presumed entrances resulting in 35 systems. All runs were performed twice with two different protocols. After the final energy minimization, MD was either started immediately without temperature control at an initial temperature of 300 K, or the cell was first gradually heated up from 10 to 300 K within 80 ps. This further stage avoided rapid displacement of the substrates away from the enzymes within the first ps of the dynamics runs. Production runs extended up to 10-14 ns. Methyl butyrate entered PLE-2 by the first method, and PLE-5 and -6 after slow heating. In PLE-3 the substrate is stuck at the entrance helices. In a further run aiming at the exchange of substrate and products, **4** entered PLE-1. This listing shows that **4** accesses the active site of PLE within a few nanoseconds. Trajectories starting from identical initial configurations, with random initial velocity distributions, are highly reproducible, but no obvious correlation between isoenzyme or protocol and the observed kinetics was found.

In two runs with PLE-1 and PLE-6, substrate (*R*)-**2** entered the active site and approached to SER 204 within 3.7 Å for 2 ns (**Fig 4**). (*S*)-**2** also passed through the entrance helices into PLE-6 and attained a similar position, but had a different orientation and the respective distance was always as high as 8-11 Å making the formation of the tetrahedral intermediate unlikely (**Fig 5**). In other runs substrates **1** and **2** remained between the entrance helices of the enzymes, and it is likely that during a longer simulation time, insertion into the active site would have taken place.

The large amide substrate **3** was placed into PLE-1 and PLE-1a and subjected to 2 ns runs. A single molecule of **3** was placed in a copy of the PLE-3 cell and run for 10ns, since this isoenzyme has the largest spacing between the entrance helices into which the large molecule might enter within the simulation time. No spontaneous access to the active site was observed, but the substrate moves between the entrance helices and would presumably enter the enzyme after a sufficiently long simulation time.

## *Tetrahedral intermediates of various substrates*

### *Stability of the tetrahedral intermediate*

For the minimizations on the eight tetrahedral intermediates, the positions of GLU 336, HIS 449, and SER 204 were fixed avoiding deformation of the presumed active site in an early stage of the calculation. In an initial molecular dynamics run at constant pressure for 20 ps the active site rearranged. In some cases GLU 336 is already replaced by GLU 452 in the active site (**Fig 6**).

The tetrahedral intermediate was then recast assuming that GLU 336 is part of the catalytic triad. Restraints with a high force constant of 28 N/m and set points around 3 Å were imposed on the distances for the atom pairs, between which proton flips occur. These are $O_\gamma$ of SER 204 to $N_\varepsilon$ HIP 449, and $N_\delta$ HIP 449 to $O_{\varepsilon 1}$ GLU 336**.** In addition to that the distances between the free $O^-$ atom of the substrate and the nitrogen atoms of GLY 126 and ALA 205 are restrained mimicking the oxyanion hole. In a further series of ten runs of 12 ps each we reduced the force constants of all restraints by a factor of two with respect to the previous one. No restraints were applied in final dynamic runs starting at 140 ps and scanning about 400 ps with substrate (*S*)-**1** in PLE-1and 5.5 to 6.5 ns for the other seven systems, respectively.

This triad was unstable, and reducing the force constants triggered exchange of GLU 336 and 452 at some stage in six out of seven runs with more than 5 ns. The distance between $O_\varepsilon$ GLU 336 and $N_\delta$ HIS 449 increased from 3 Å to about 10 Å, and the distance of $O_\varepsilon$ GLU 452 to $N_\delta$ HIS 449 decreased in turn to 2.7 Å (**Fig 6**). In the remaining system the active site had been deformed by the restraints, and the two GLU residues 336 and 452 could not relax during the simulation time.

*Substrate pockets*

Amino acids of the respective enzyme with van der Waals contacts to the substrate were assigned to binding pockets (**Tab 1**). In addition to seven tetrahedral intermediates with more than 5 ns run time, four Michaelis-Menten complexes were analyzed. They were generated by pulling substrate (*R/S*)-**1** through the entrance helices into the active site of PLE-1 and by spontaneous access of substrates (*R/S*)-**2** into PLE-6. The mechanism of generating the pockets is different for these systems from that for the tetrahedral intermediate. Pulling the substrate might deform its environment and during spontaneous access the substrate might not fully insert into the active site.

A total of 33 residues have contact to the substrate in at least one of the eleven systems and are involved in the formation of pockets (**Tab 1**). 17 (VAL, LEU, MET, ILE, PHE) are highly and 13 (THR, ALA, TYR, SER, GLY, HIS) are moderately hydrophobic, respectively [36]. Only three charged amino acids (GLU) are participating in the pockets. Their composition varies strongly, and only eleven amino acids are found close to the substrates in at least five of the eleven systems. The amino acids are part of nine different elements of the secondary structure. By aligning PLE to hCE it was shown, that eleven of the 33 pocket residues in our runs have analogs among the 14 residues forming the hydrophobic pockets around naloxone methiodide in ref. [14a]. This large substrate has more than twice the number of amino acid neighbors than the phenyl acetates showing that the substrate size has significant influence on the shape of the active site. Nevertheless a significant correlation exists between the amino acids participating in the respective pockets as generated by four different approaches (predefined tetrahedral intermediate, Michaelis-Menten complexes generated by restrained and spontaneous access, and X-ray diffraction experiments [14a] (**Tab 1**).

The substrates of our tetrahedral intermediates had arbitrary orientations after manual positioning in the active site, but the initial configurations were not trapped in local energy minima. Comparison of the pattern of amino acids adjacent to substrate (*R*)-**1** in PLE-1 at the beginning and at the end of the 6 ns run reveals strong rearranging of the enzyme around the substrate (**Fig 7**), and phenyl ring and acyl rest exchange their positions. The amino acids, in which the phenyl ring is finally embedded, are wide spread in the enzyme at the beginning. Plots of characteristic atomic distances between the substrate and surrounding amino acids indicate that the relaxation took place within the first 1-2 ns of the simulation. Consequently we consider the seven systems, which were subjected to 6 ns runs, as fully relaxed. The stable final position is not biased by manual positioning of the substrate.

The pockets are very flexible and adapt to substrates with different sizes and to both enantiomers (induced fit, cf. [14a, 16]). This flexibility may be due to the fact that a large number of hydrophobic residues on several secondary structure elements inside the protein may reorient and approach to the substrate. It is difficult to compare our pockets with those of the rigid Toone model and to find modifications induced by mutations. In our simulations the enantiomers may fit about equally well and it is difficult to find indications of enantioselectivity.

Comparison of the positions of the amino acids in the series of isoenzymes confirms this picture. Evaluation of root mean square displacements (RMSD) of pocket residues for PLE-2 to 6 and APLE with respect to PLE-1 yielded strongly different values making it impossible to identify a rigid and a flexible pocket. The second entrance helix (338-346), which is part of both acyl and phenyl pockets, has very different positions in the series of isoenzymes and RMSD values up to 3-6 Å. On the other hand the position of the random coil 124-128 is very well conserved with RMSD values mostly around 1-2 Å even though the GLY residues might result in a flexible structure. The carbonyl oxygen of GLY 124 interacts with the amide proton of LEU 127 and stabilizes this coil.

### *Hydrophobic pockets*

According to Toone [17] two hydrophobic pockets $H_L$ and $H_S$ of different shapes accommodate the phenyl and the alkyl rests of the ester, respectively. The clear distinction between rigid alkyl and flexible phenyl pockets as in [14a] is not possible here. Both pockets strongly overlap, and identical residues, as e.g. LEU 338 are part of both in some cases (**Tab 1**). Amino acids of all nine secondary structure elements arrange around the phenyl ring and the acyl rest. Comparing different systems such as (*S*)-**1** and (*R*)-**2** in PLE-1a, residues such as PHE 297 may be part of the phenyl in the one and of the acyl pocket in the other tetrahedral intermediate.

### *Oxyanion hole*

Typically, amidic protons in the backbone stabilize the carboxyl oxygen of the ester substrate, and by steric reasons glycine residues are favored there. From the respective distances for substrates **1** and **2** to backbone N-H, the oxyanion hole was assigned to a random coil below one of the entrance helices. This coil contains the GGGX motive (124-127) which is related to capacity of PLE to convert both secondary and tertiary alcohols [1, 37]. In PLE the ester C in the tetrahedral intermediate is considered to have (*S*)-conformation, and the free oxyanion can be stabilized by GLY 125, 126 and LEU 127. Apart of the glycine rich random coil only ALA 205 on the helix around SER 204 participates in the formation of the oxyanion hole (**Tab 1**). For substrates (*S*)-**2** in PLE-1 and (*R*)-**1** in PLE-1a no stabilization of the oxyanion was found. Interestingly, these two combinations of substrate and enzyme show extremely efficient enantioselective catalytic reaction [7].

*Acetyl pocket*

The ethyl- and especially the methyl groups are too small for inducing formation of alkyl pockets and filling them, but in contrast to earlier work we find well defined pockets around the acetyl group of the substrates. All secondary structure elements around the substrate with the exception of the random coil containing PHE 297 are found as part of theses pockets in varying composition (**Tab 1**). The acetyl group is related to the oxyanion and consequently the random coil 124-135 plays a major role, but also the helices containing the active site residues SER 204 and HIS 449 are strongly represented.

## *Electron transfer reaction: QM-calculations*

By first principles molecular dynamics the proton transfer and splitting of the substrates (*R/S*)-**1** was simulated. The starting structures were cut out from those of the tetrahedral intermediates in PLE-1. The model contained the side chains of the catalytic triad (SER, GLU and doubly protonated HIS) including the respective $C_\alpha$-atoms, which were fixed and capped off by three hydrogen atoms each. The geometry of the active site was stabilized by constraining the two distances between the hydrogen which flipped from serine $O_\gamma$ to histidine $N_\varepsilon$ to these atoms as well as the bond between the oxygen of the serine and the ester C1 atom of the (*R/S*)-**1** substrate. The environment of the three amino acids of the active site and of the covalent bound substrate was filled up by 13 molecules of water which form a cluster. The oxyanion of the substrate, which is normally stabilized by hydrogen bonds to the oxyanion hole, was capped with a hydrogen atom.

The decay of the tetrahedral intermediate of the substrate (*R*)-**1** was initiated by gradually increasing the constrained distance between substrate O and capping H. This results in a highly polar ester bond. The negatively charged O2 is saturated by capturing a proton from an $H_2O$ molecule. Splitting of the ester bond occurs within 760 fs (**Fig 8**) and the proton at the histidine $N_\varepsilon$ flips to the alcohol. The acid remains bonded to SER 204, and the alcohol is set free from the enzyme, but remains connected to the water cluster. As this approach principally cannot reproduce enantioselectivity, the system with the (*S*)-enantiomer was not further studied.

## *QM/MM simulation of the full system*

The QM calculation on the active site was enhanced by an MM description of the whole enzyme reproducing the shapes of the active site and of the binding pockets. The enzymatic reaction with methyl butyrate was reproduced starting with the final configuration of the force field simulation of spontaneous access to the active site with a final C1-$O_\gamma$ distance of 6 Å. By a variable constraint this distance was reduced to 3 Å within 0.5 ps of the QM/MM run. As a result of the short distance the system generates a covalent bond forming the tetrahedral intermediate.

The complete substrate and the side chains of the catalytic triad consisting of SER 204, HIS 449 and GLU 452 are modeled quantum mechanically starting at the respective $C_\beta$-atoms. The water box used in the force field simulation was omitted here since no significant reaction of the enzyme to the missing dielectric is to be expected during the short simulation time of about 4 ps, but four QM water molecules were provided for assisting the reaction. The QM cell has a total charge of zero after adding one $Na^+$ counter ion.

Characteristic distances in the actives site are plotted as a function of time (**Fig 9**). Reaction was initiated by reducing the constrained distance between the ester C1 and the $O_\gamma$ of SER 204 until a covalent bond was formed after 1 ps (*blue*), being stable during 2.8 ps. The enzymatic splitting of the ester is indicated by the black line which indicates an increase of the ester bond from 1.4 to 3-4 Å within 0.2 ps after forming the tetrahedral intermediate. This reaction was so fast in our QM/MM calculations that no distinction between the kinetics of different substrates and enantiomers seems to be possible at this point.

The red and the green lines indicate proton transfers accompanying the ester cleavage. For the formation of the tetrahedral intermediate the transfer of $H_\gamma$ from the side chain of SER 204 to $N_\varepsilon$ of HIS 449 is essential. In the starting configuration the HIS side chain had an unfavorable orientation and the transfer had to incited by a second constraint (*green*, between 0.5 and 1 ps simulation time). During the ester splitting a methanol is formed which is extremely basic and spontaneous transfer of the H from the $N_\varepsilon$ of the now doubly protonated HIS 449 to this methanolate anion is seen at about 1.2 ps simulation time (*red*).

Bond break between the remaining acid and the serine is induced by approach of a water molecule to this bond. One water proton adds to the carboxyl oxygen bonded to the serine, and the remaining OH-group adds to $C_\beta$ of SER 204, restituting its full side chain.

### *Release of the products from the enzyme*

In an unrestrained force field simulation eight molecules of substrate **4** were positioned close to PLE-1 and PLE 1a, and one butyric acid and one methanol molecules were put into the active sites as products of splitting of this substrate. No exchange of educts and products was observed for PLE-1 within 32 ns. One substrate enters PLE-1a within 0.3 ns (**Fig 10**), and the distance between $O_\gamma$ of SER 204 and the ester C4 reduces from 20 Å to about 3.2 Å. For methanol the C1-$O_\gamma$ distance is about 12 Å for 25 ns, until the molecule spontaneously leaves the active site. Butyric acid is stable inside the active site with a distance of 2.7-3 Å between its carboxyl C4 and the side chain oxygen of serine.

# 4 Conclusion

The full reaction mechanism of ester hydrolysis by PLE is consistently described validating our force field and quantum mechanical models, and our previous assignment of GLU 452 instead of 336 to the catalytic triad is confirmed in calculations with substrates.

We suggest a soft model of the system enzyme –substrate providing no rigid enzyme structure and discarding simple lock and key mechanisms. Motion of a substrate molecule through the enzyme is described as a combination of flow in viscous oil and jump diffusion with barriers around 10 $k_BT$ for phenyl acetate substrates. Several channels for the access to the hidden active site may be possible in such a structure, but in our calculations and may be also in reality only very short ones with low barriers seems to play a role. Before entering the enzyme the substrate may stick to the surface or even entangle with surface structures. The protein structure relaxes after perturbation of the entrance channel and active site by the substrate within times of the order of 1 ns being very fast as compared to typical turnover numbers of PLE [23], but slow with respect to the enzymatic reaction itself. The local relaxation involves a series of mainly hydrophobic amino acids and results in an induced fit around the substrate.

Unrestrained spontaneous access to the active site is in principle fast enough for force field simulations and leads to Michaelis-Menten complexes. The small substrate seems to enter the enzyme without significant barrier, but for the phenyl acetates access was slow, and for PBNA it could not be seen in the time window of the simulation. We thus speculate that the access is thermally activated, and that barrier heights in the entrance site have influence on the enzyme kinetics. These barriers depend on the shape on the substrates.

By classical MD the binding pockets around the active centre were identified. Spontaneous access does not always result in very small distances between ester carbon and SER $O_\gamma$, and enantioselectivity may be a consequence of slightly smaller or bigger distances of the respective ester C to the serine $O_\gamma$. Steric preference of one enantiomer inside the active site is possibly not rate determining. A typical key-lock effect in the active site would manifest itself by noticeable differences between the stabilization of the two enantiomers in the pockets, which is not seen.

The actual reaction in the active site includes very fast proton transfers on a ps scale and is probably not rate determining. Astonishingly the release of the reaction products seems to be very slow and took about 20 ns even for the smallest substrate which enters the active site within less than one ns. For larger substrates unrestrained product release seems to be beyond reasonable computing time even though entropy should favor release of products with respect to uptake of educts.

# Acknowledgments

We acknowledge assistance of Dr. R. Kourist, U. Stiehm, D. Wagner and C. Ziegler (Greifswald) during some of the calculations, and fruitful discussions with Dr. E. Brüsehaber and A. Hummel (Greifswald).

# Figure Captions

**Fig 1:** Structural formulae and atom names for substrate (**1-4**) and products (**5, 6**). The phenyl-alkyl-acetates **1** and **2** are identical to substrates **1** and **6** in [7].

**Fig 2:** Access of (*S*)-**1** into PLE-1by restrained MD. Enzyme coloring indicates the α-helices (*red*) and β-strands (*orange*) of the rigid hydrolase fold, the entrance helices (*purple*) and further flexible helices and coils (*green*). The active site is visible in part and colored according to the elements. Substrate snapshots are painted as a function of simulation time (total 1.8 ns) as indicated by the color bar on the left. It is clearly seen that the substrate passes straight through the entrance helices.

**Fig 3:** Force distance curves of (*R*)-**1** entering PLE-1 at temperatures of 300 (initial), 300, 400, 500 and 600 K and a pulling velocity of 2 m/s (*from bottom*). Fluctuations on a time scale below 10 ps or 0.6 Å are smoothed. A triangular barrier with a height of 0.45 nN and a length of 7 Å corresponding to about 1 eV is indicated in the bottom trace. Simulation is too short for observing denaturation at high temperatures, but softening and decrease of barrier heights is clearly seen.

**Fig 4:** Plot of the distance between $O_\gamma$ SER 204 and the ester C3 atom of the substrate over the time indicating spontaneous access of substrate (*R*)-**2** into PLE-6. For the starting position (painted *red* in the insert) the distance is about 18 Å. Access to the active site (*colored by elements*) takes less than 200 ps, and around 2 ns a position of the substrate with a distance of about 3.7 Å is stable for more than 1 ns (*light brown*). After 4 ns a very stable position with a distance of about 7 Å is attained (*light green, blue, dark blue*). In the distance plot, the colors of corresponding substrate snapshots in the insert are indicated.

**Fig 5:** Plot of the final positions of substrates (*R*)-**2** (*left*) and (*S*)-**2** (*right*) relative to the active site after spontaneous access (all colored *by elements*). The overall positions of the two enantiomers relative to the entrance helices (*purple)* are similar, but due to the different orientations, the distances between $O_\gamma$ SER 204 and the ester C3 atom are different (7 vs. 11 Å).

**Fig 6:** Plot of atomic distances over time of the tetrahedral intermediate of (*R*)-**1** in PLE -1 indicating exchange of GLU 336 by GLU 452 in the catalytic triad. Distances between the two O-atoms of the GLU side chains are indicated in *green* and *blue* (336), and in *black* and *red* (452), respectively. Traces for the first 20 ps reflect an unrestrained **NpT** run. During this time GLU 452 approaches to HIS 449, and GLU 336 leaves its original position. Restraints imposed after 20 ps result in restitution of the original triad with GLU 336. After about 105 ps of simulation time the force constants of the restraints are down to 0.2 N/m, and GLU 452 flips back into the active site. Finally at 200 ps GLU 336 moves even further out attaining a distance of about 10 Å to the HIS.

**Fig 7:** Soft substrate environment and induced fit. The figure shows the phenyl pockets for substrate (*R*)-**1** (*colored by elements*) in PLE-1 at the beginning (*left*) and end (*right*) of a 5.5 ns free MD run. Amino acids forming the pocket at the beginning are painted *red* in both views; those hosting the phenyl group at the end are *green*. It is clearly seen that the ring reorients during the run and only has contact to the *red* amino acids at the beginning and to the *green* ones at the end. During the run, the *green* amino acids rearrange from a lengthy structure to a shell around the phenyl, whereas the *red* pocket melts down to globular structure.

**Fig 8:** Fast ester splitting of (*R*)-**1** in a QM calculation. Reaction is triggered after 400 fs by increasing the capping proton from the carboxyl O2 of the ester (*black line*). The bond of the ester O2 to the phenyl C3 (*green*) remains stable, but the second C1-O1 bond to the methyl group breaks (*blue*), and the atom-atom distance increases to 1.8 Å. The resulting negatively charged O1-atom attracts a proton from a water molecule and the O1-H-distance decreases to 1.2 Å (*red*).

**Fig 9:** Atomic distances over time for the QM/MM calculation of the splitting of methyl butyrate (**4**) in PLE-1a. Reaction is induced by pulling simultaneously the $H_\gamma$ of SER 204 towards $N_\varepsilon$ of HIS (*green line*) and in turn the ester C1 of the substrate towards the $O_\gamma$ of the SER (*blue*). The ester splitting is seen as an increase of the distance between ester C1 and O1 after 1.2 ps (*black*). The methanolat anion, which is set free during this step, is very basic and attracts the former SER proton (*red*), which intermediately was connected to $N_\varepsilon$ HIS.
The bond between ester C1 and SER $O_\gamma$ is broken by inserting a water molecule. Pulling the oxygen of this molecule to the carboxyl C1 only was not sufficient for forming and releasing the butyric acid molecule (*magenta, between 2 and 2.8 ps*). A four center transition state had to be generated by pulling one of the water protons close to the $O_\gamma$ of SER (*light blue*) simultaneously to shortening the water O–carboxyl C1 distance. In this approach, the bond between acid and enzyme (*blue*) breaks after 3.8 ps.

**Fig 10:** Release of methanol from the splitting of **4** in PLE-1a. Distances between $O_\gamma$ SER 204 and carbon atoms of three molecules in the active site are plotted as a function of time. These atoms are C4 of butyric acid (*black*), C1 of methanol (*red*), and C4 of the one of the eight educt molecules around the substrate, which enters the active site spontaneously at the beginning of the run. The increase in the black line after 25 ns indicates that the methanol is ejected from the enzyme. In the insert the path of the methanol molecule is visualized by coloring it according the simulation time (cf. reference bar at the left). The enzyme is painted *cyan*, except of the entrance helices (*purple*) and of the catalytic SER (*by elements*).

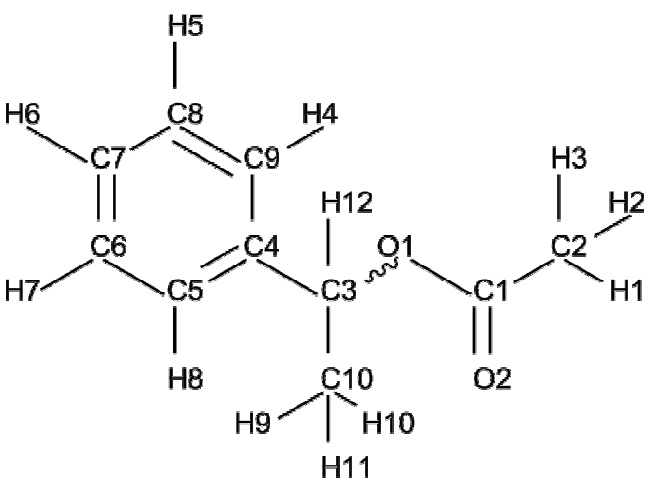

(*R/S*)-1-phenyl-1-ethylacetate (**1**)

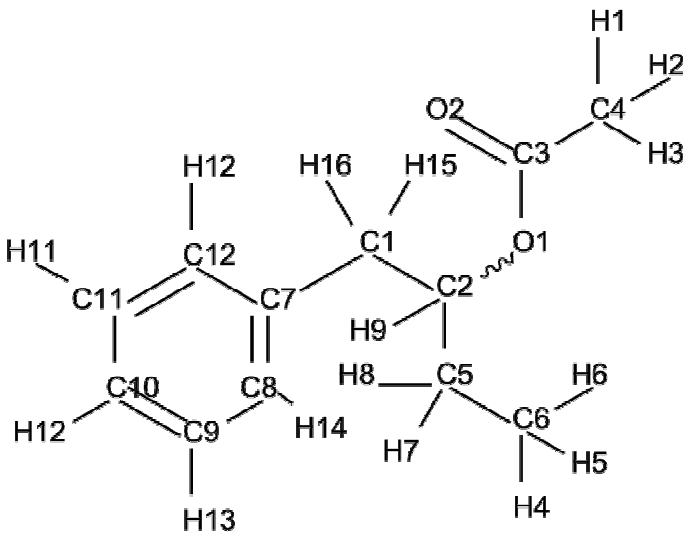

(*R/S*)-1-phenyl-2-butylacetate (**2**)

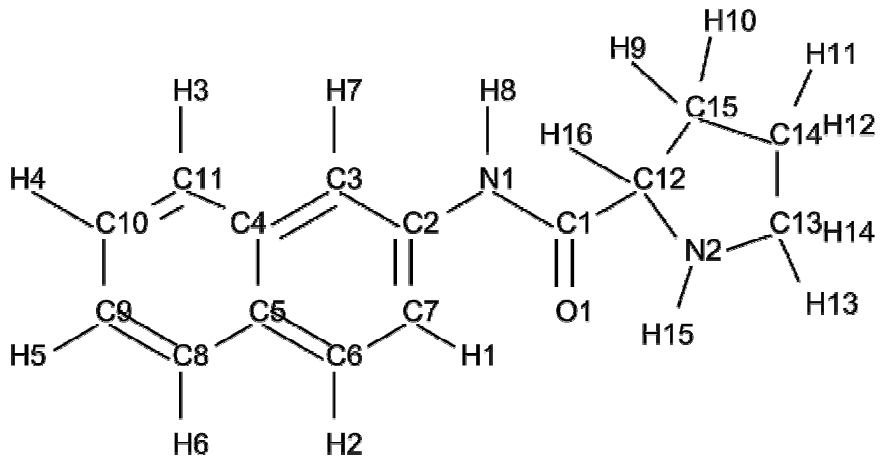

proline-β-naphthylamide (**3**)

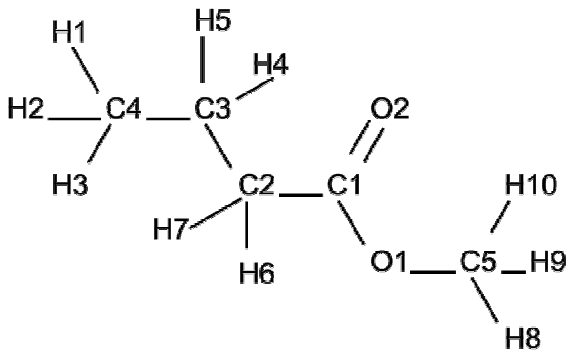

methylbutyrate (**4**)

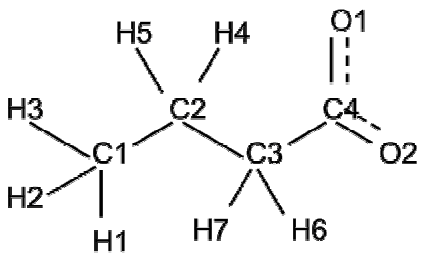

butyric acid (**5**)

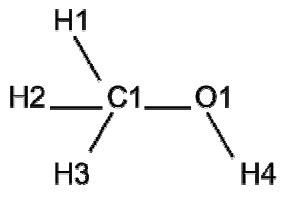

methanol (**6**)

# Fig 1

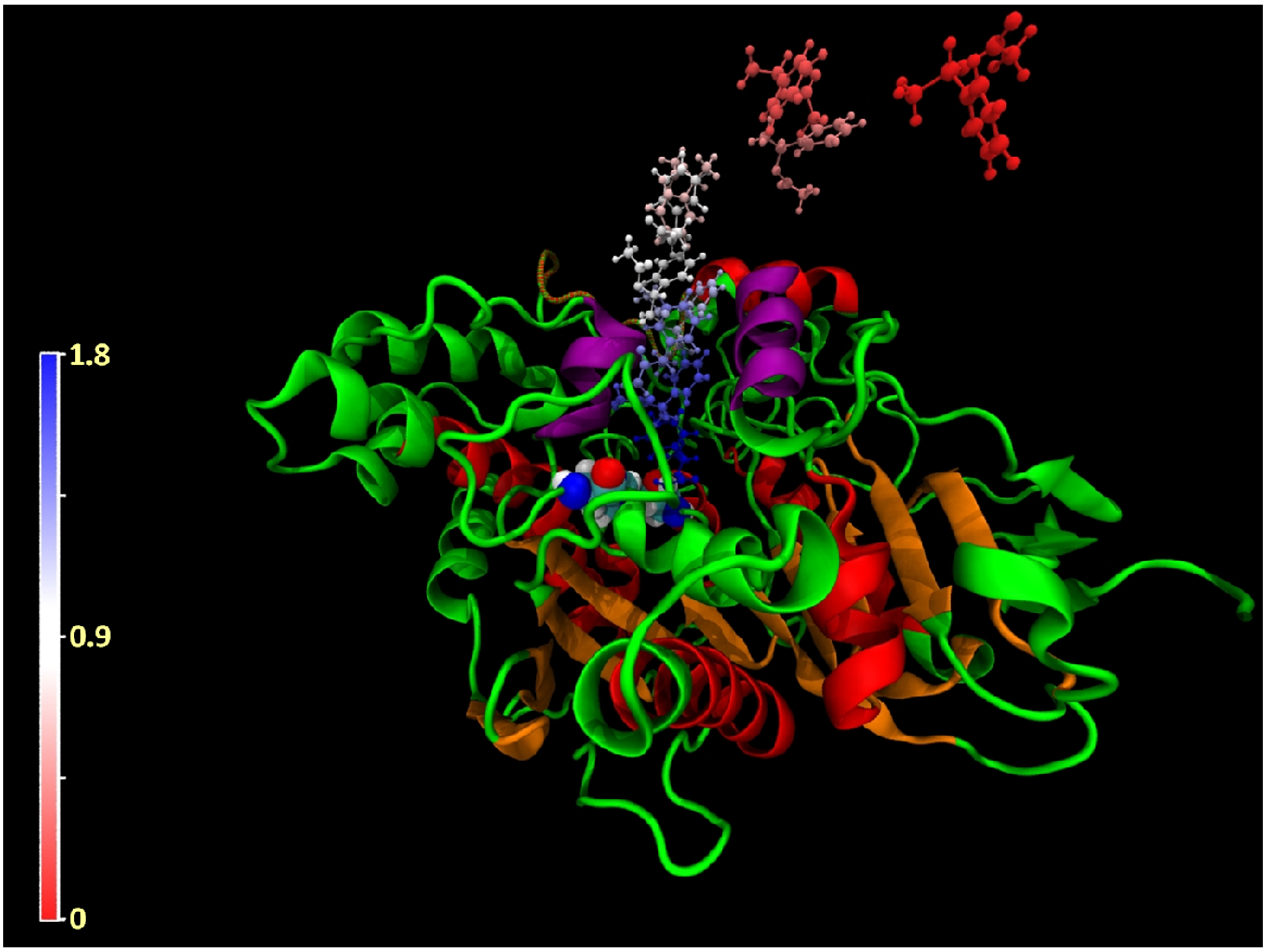

# Fig 2

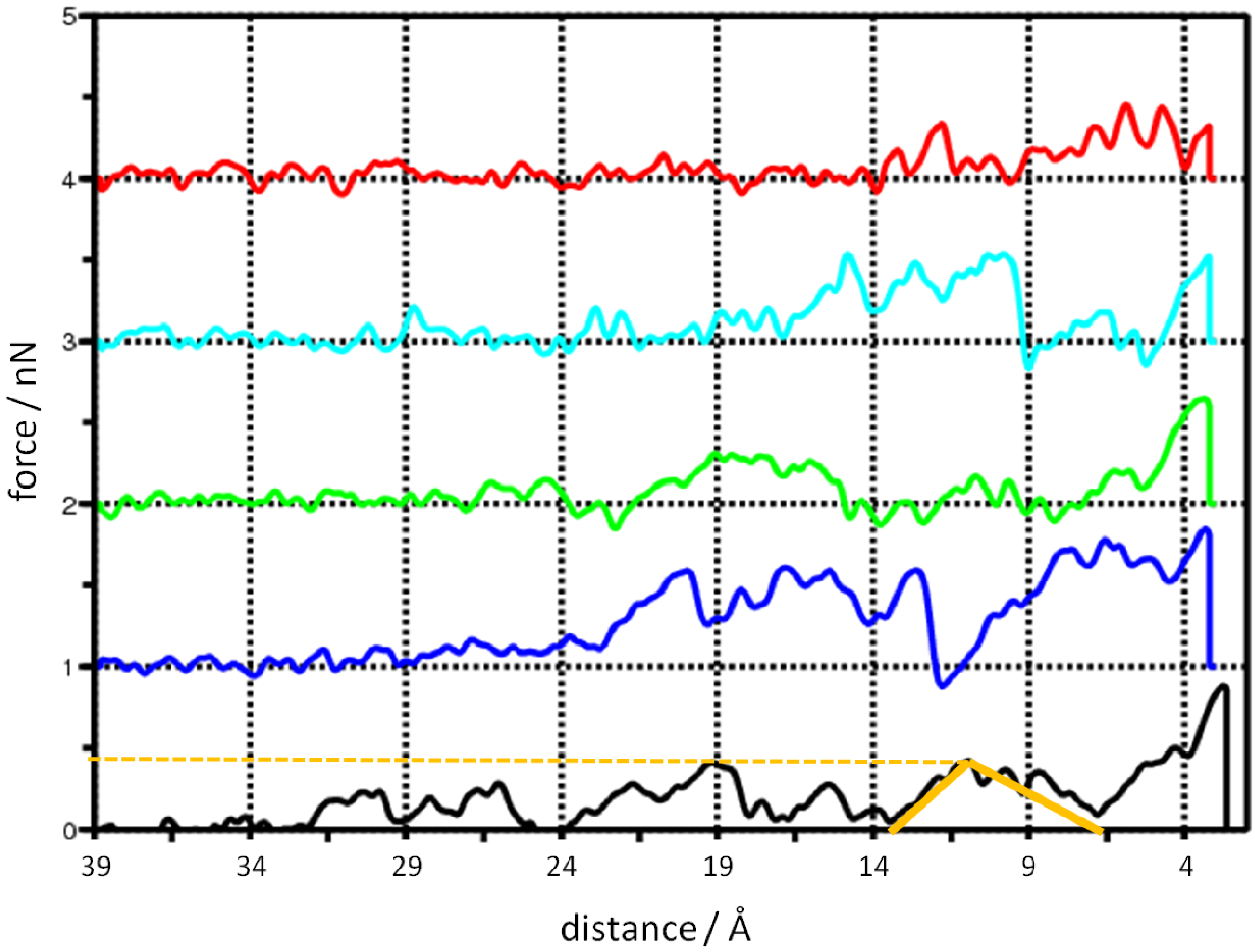

**Fig 3**

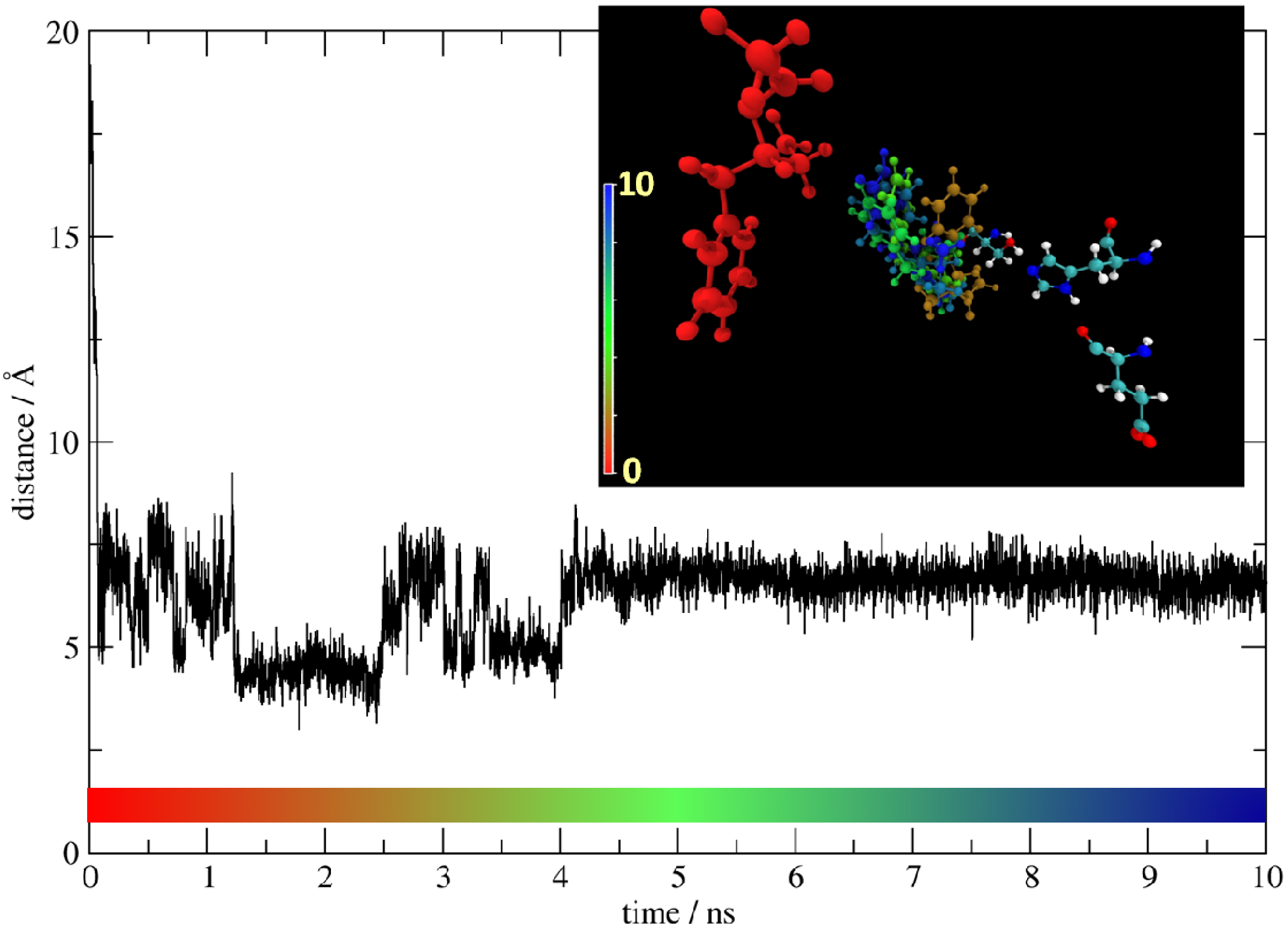

**Fig 4**

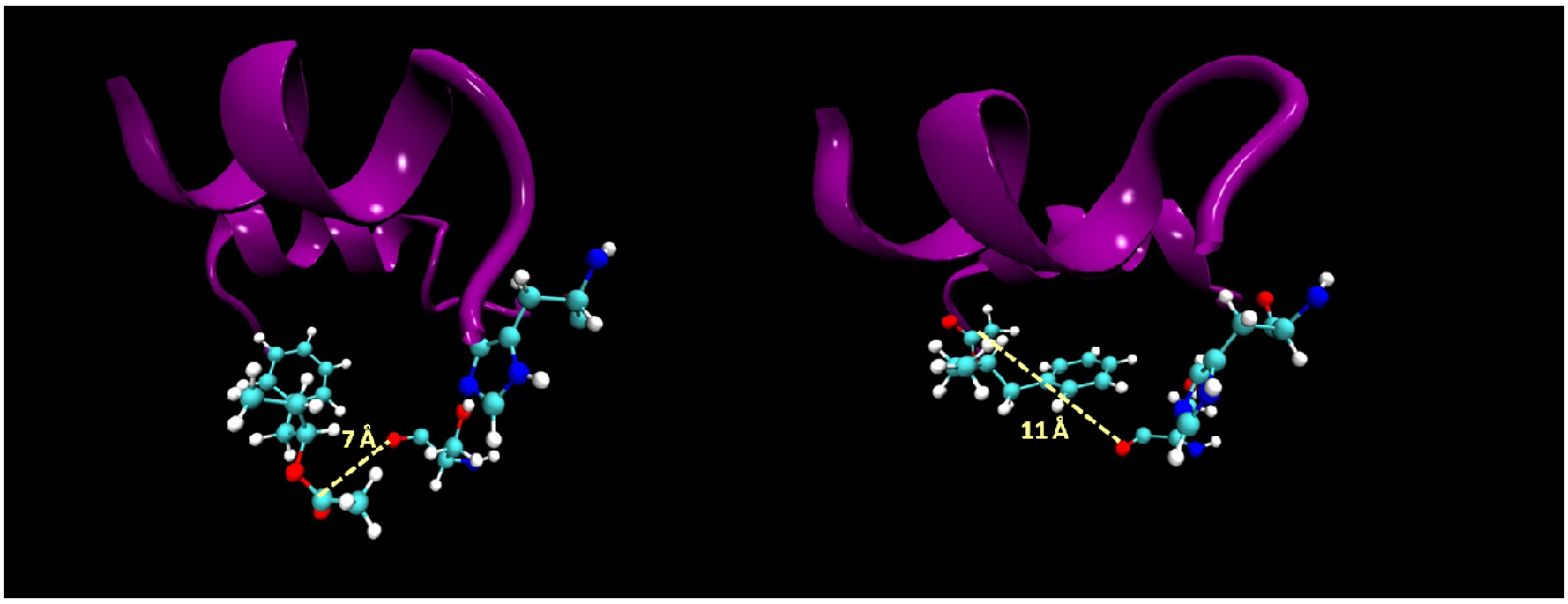

**Fig 5**

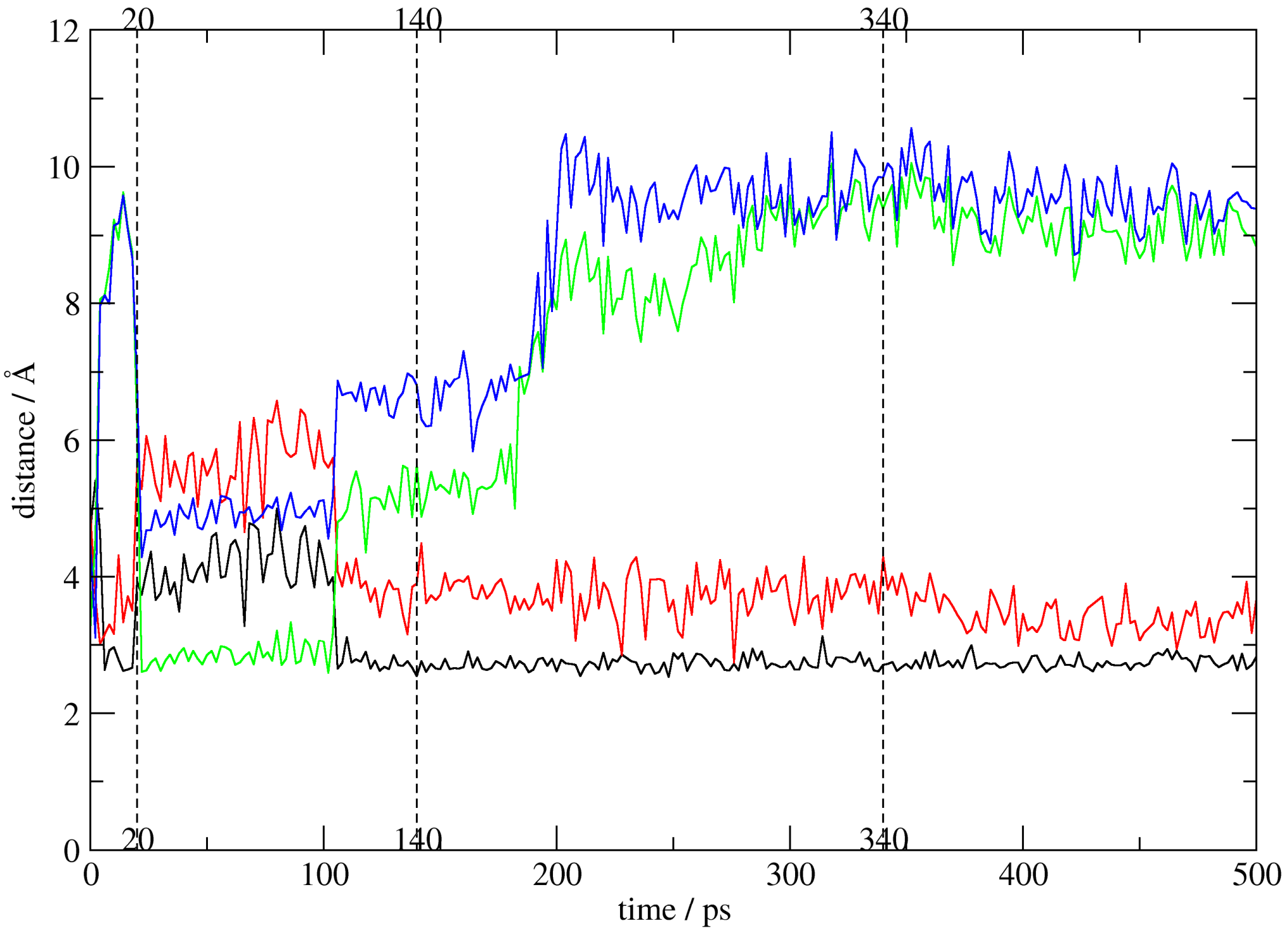

# Fig 6

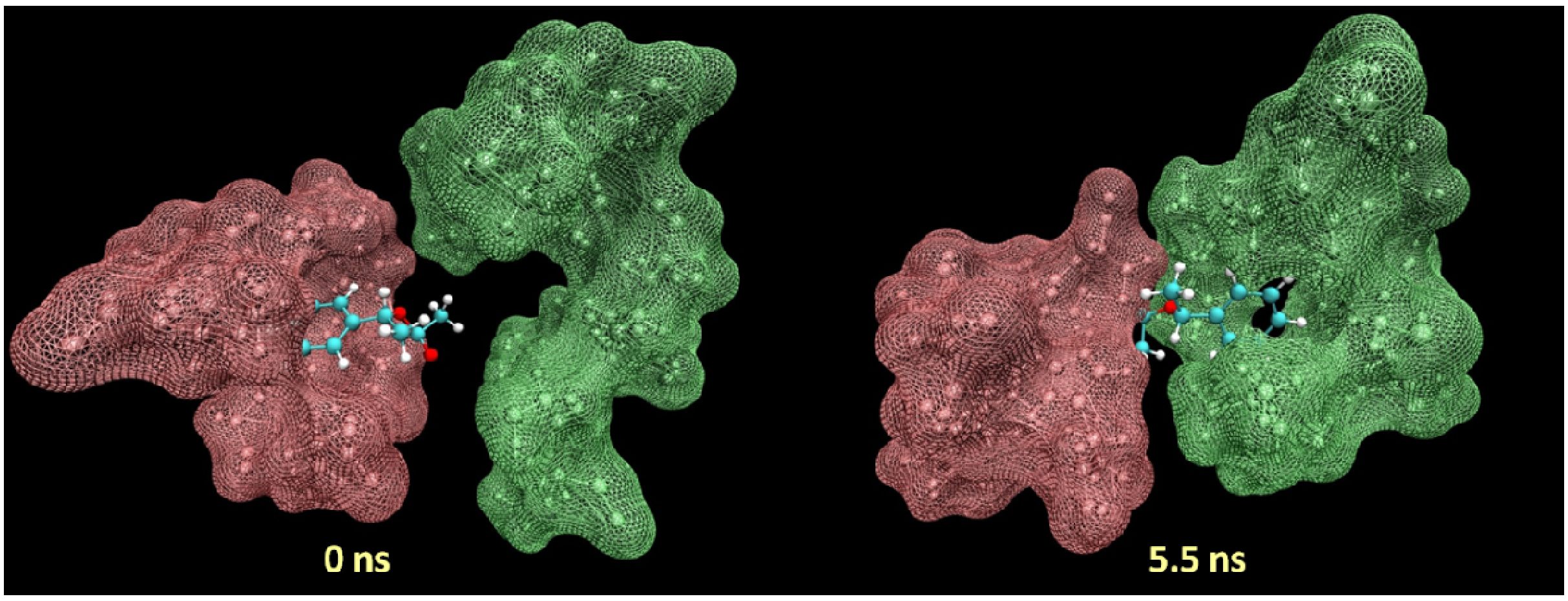

0 ns
5.5 ns

# Fig 7

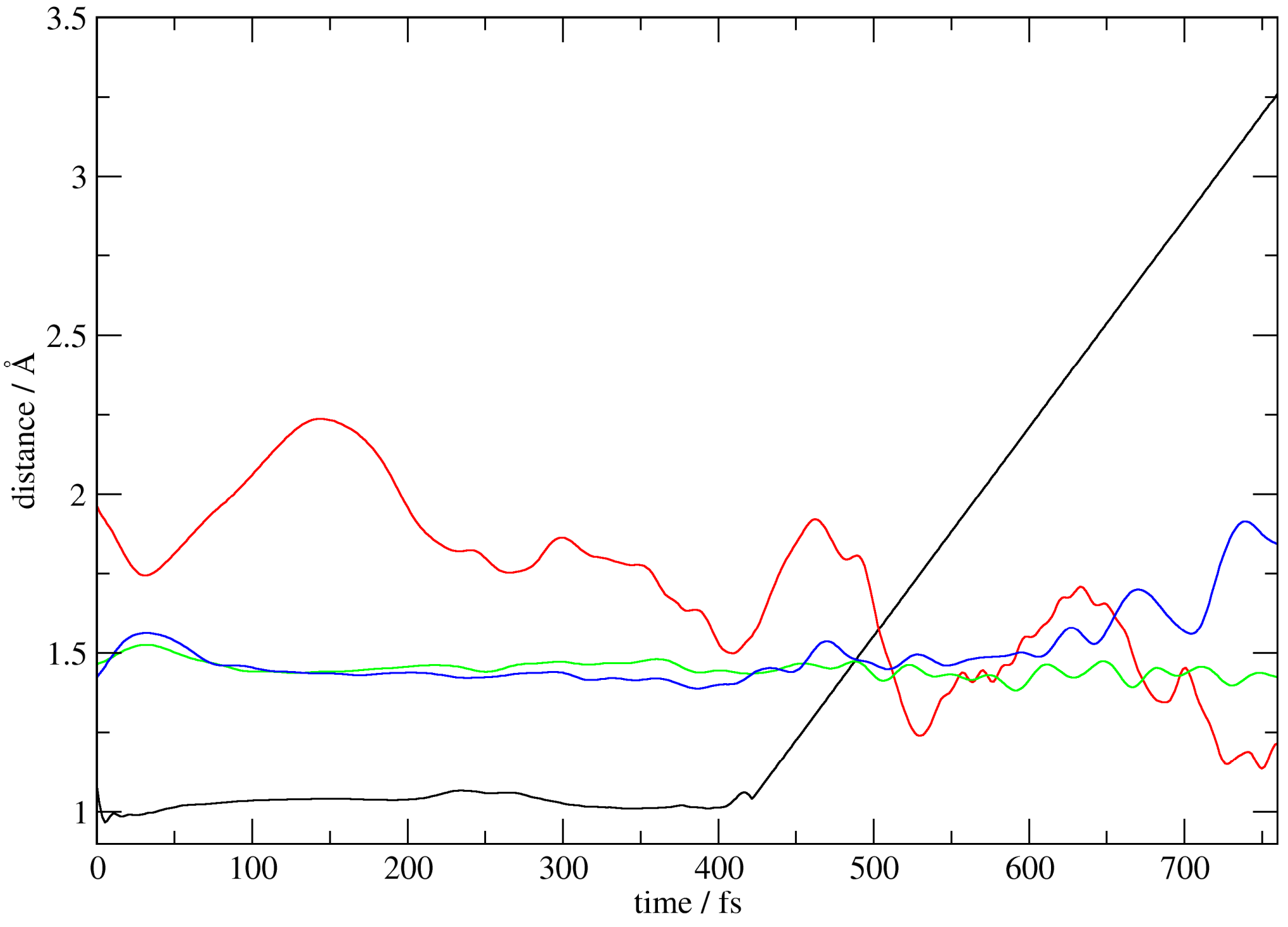

**Fig 8**

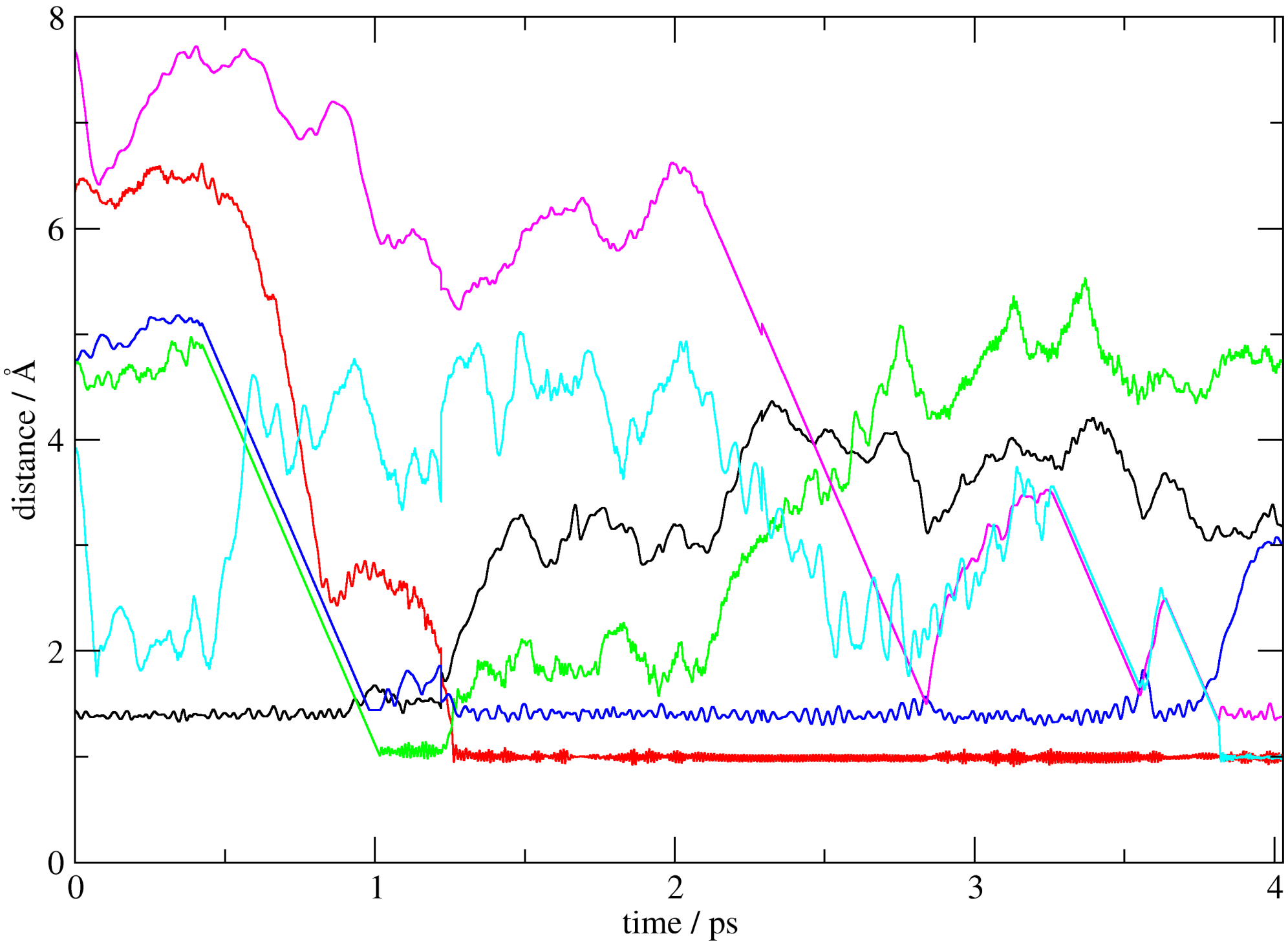

**Fig 9**

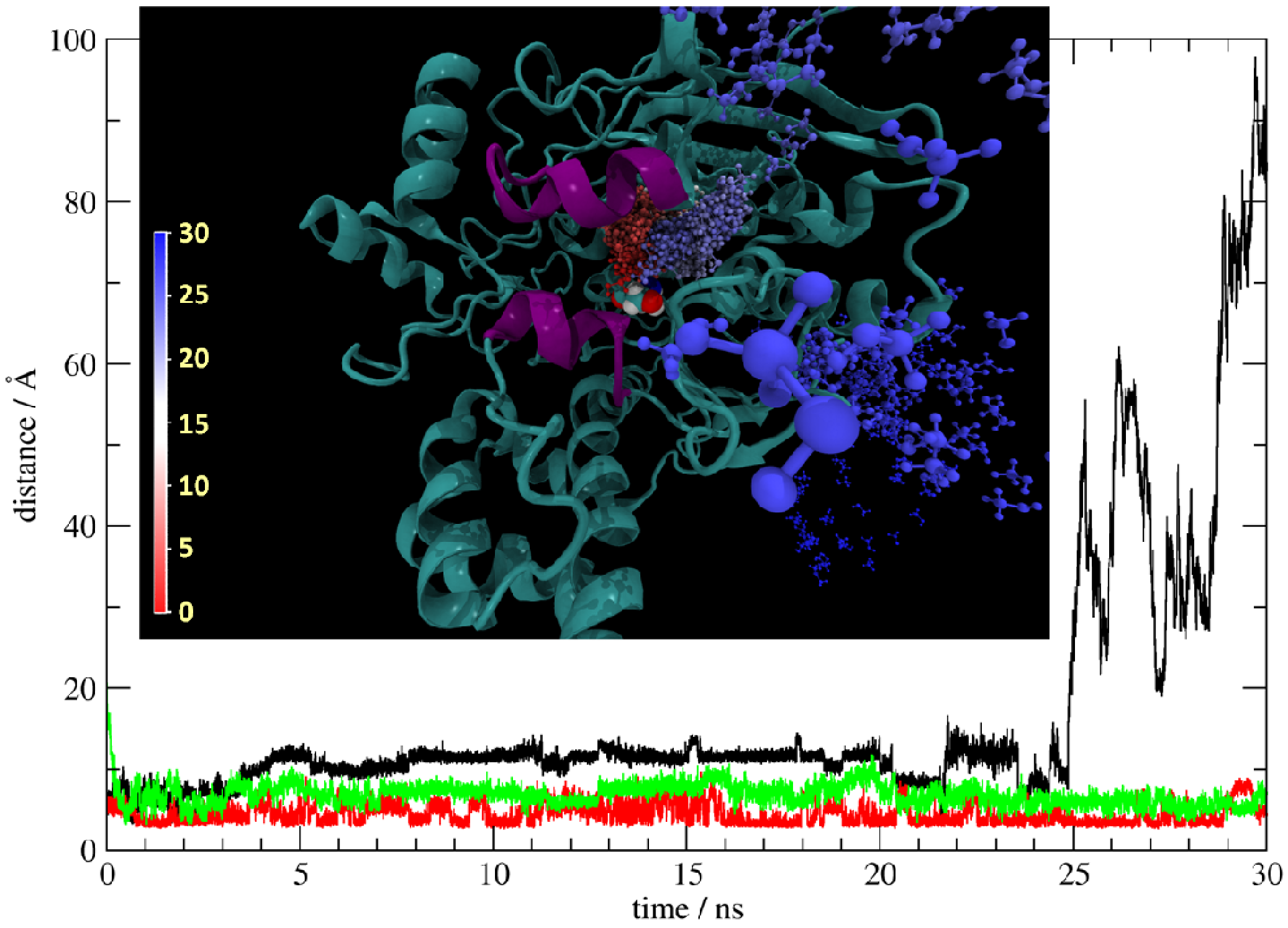

**Fig 10**

**Tab 1:** Amino acids participating in the respective binding pockets of tetrahedral intermediates and Michaelis-Menten complexes. The respective enzymes and substrates are shown in the second and third lines, respectively. The respective pockets around each substrate are indicated by X (phenyl), O (alkyl), Δ (acetyl), and W (oxyanion hole). Several residues are part of more than one pocket. The second column lists the analogs in the hydrophobic pockets around naloxone methiodide in hCE [1]Ref. [14a]. The residues in PLE-1 are regrouped on nine secondary structure elements (third column).

| PLE-1 residue | hCE[1] residue | secondary structure | tetrahedral intermediate | | | | | | | | access by restraints | | spontaneous access | |
|---|---|---|---|---|---|---|---|---|---|---|---|---|---|---|
| | | | PLE-1 | | | | PLE-1a | | | | PLE-1 | | PLE-6 | |
| | | | (*R*)-**1** | (*R*)-**1** | (*R*)-**6** | (*S*)-**6** | (*R*)-**1** | (*S*)-**1** | (*R*)-**6** | (*S*)-**6** | (*R*)-**1** | (*S*)-**1** | (*R*)-**6** | (*S*)-**6** |
| MET79 | LEU96 | entrance | 0ns | <6ns | | | | | | | Δ | X | X | Δ |
| LEU83 | LEU100 | helix | | | | X | | | X | | O | X | | |
| PHE84 | PHE101 | 72 – 87 | O | Δ | | X | | | X | | X | X | X | X |
| GLY124 | | oxyanion hole | | | | | | Δ | | XW | | O | | |
| GLY125 | | random coil | O | Δ | | X | X | ΔW | Δ | | X | O | | |
| GLY126 | | 124 – 127 | | XΔW | Δ | Δ | X | OW | ΔW | XΔ | | | X | X |
| LEU127 | | | | XΔW | OΔW | Δ | O | | OΔW | X | | | X | Δ |
| VAL128 | | | | | X | | | | | | | | X | Δ |
| TYR135 | | | | | | X | | | | | X | X | | |
| GLU203 | | active site, | | Δ | Δ | X | | Δ | Δ | | | O | | |
| SER204 | | followed by turn | X | OΔ | OΔ | Δ | Δ | Δ | Δ | XΔ | | | | X |
| ALA205 | | and $3_{10}$-helix | Δ | | | | | ΔW | ΔW | XW | | O | X | X |
| GLY207 | | | | | | | | | | X | | | | |
| GLU208 | | | | | O | | | | | X | | | | |
| SER230 | | random coil | X | | | X | Δ | | | | | | | |
| GLY231 | | | | O | | | | X | | | | | Δ | |
| VAL232 | | | | | | O | | | | | | | Δ | |
| ALA233 | | | | O | O | | Δ | | | X | | | Δ | |
| THR235 | THR252 | | | | | | | | | X | | | | |
| LEU238 | LEU255 | | | X | | | | | | X | | | O | O |
| LEU286 | LEU304 | $\alpha$-helix, turn | | X | X | Δ | X | | X | | Δ | | X | O |
| PHE300 | LEU318 | end of turn | | X | X | | O | O | X | | | | X | |
| GLU336 | | second | | | | | | X | | | | | | |
| LEU340 | | entrance helix | | Δ | XΔ | XO | X | X | X | OΔ | O | X | | X |
| LEU341 | LEU358 | 338 – 346 | Δ | X | XΔ | OΔ | XO | X | X | O | | | X | X |
| MET344 | LEU363 | | | X | X | XOΔ | X | X | X | | O | X | | X |
| MET345 | MET364 | | | X | X | | O | | X | | | | O | O |
| VAL406 | MET425 | $\alpha$-helix | | | | | | | O | | | | O | |
| PHE407 | PHE426 | | Δ | O | O | | OΔ | XO | O | O | | | Δ | O |
| HIS449 | | random coil | XΔ | OΔ | Δ | XO | | XΔ | Δ | OΔ | | Δ | | X |
| GLY450 | | | X | Δ | | X | | | Δ | Δ | | Δ | | |
| ILE453 | | | X | | | X | | | Δ | | | Δ | | |
| PHE454 | | | | | | X | | | | | | | | |